\documentclass[twocolumn,superscriptaddress,aps,prb]{revtex4-1}
\usepackage{graphicx}
\usepackage{amsmath}
\DeclareMathOperator{\sgn}{sgn}

\begin{document}
\title{Edge states in dichalcogenide nanoribbons and triangular quantum dots}
\author{C. Segarra}
\affiliation{Departament de Qu\'{\i}mica F\'{\i}sica i Anal\'{\i}tica, Universitat Jaume I (UJI), Av.\ de Vicent Sos Baynat, s/n, 12071 Castell\'o de la Plana, Spain}
\affiliation{Department of Physics and Astronomy and Nanoscale and Quantum Phenomena Institute, Ohio University, Athens, Ohio 45701-2979, USA}
\author{J. Planelles}
\affiliation{Departament de Qu\'{\i}mica F\'{\i}sica i Anal\'{\i}tica, Universitat Jaume I (UJI), Av.\ de Vicent Sos Baynat, s/n, 12071 Castell\'o de la Plana, Spain}
\author{S. E. Ulloa}
\affiliation{Department of Physics and Astronomy and Nanoscale and Quantum Phenomena Institute, Ohio University, Athens, Ohio 45701-2979, USA}
\email{}
\date{\today}


\begin{abstract}
The electronic structure of monolayer MoS$_2$ nanoribbons and quantum dots have been investigated by means of an effective $\mathbf{k}\cdot \mathbf{p}$ two-band model.
Both systems with borders exhibit states spatially localized on the edges and with energies lying in the band gap. We show that the conduction and valence band curvatures determine the presence/absence of these states whose origin has been related to the marginal topological properties of the MoS$_2$ single-valley Hamiltonian.
\end{abstract}

\maketitle

\section{Introduction} \label{sec:intro}

Since the discovery of graphene,\cite{Novoselov2004} atomically thin layered structures have attracted growing interest and several
new two-dimensional (2D) materials have been prepared, \cite{Novoselov:2005aa} including hexagonal BN and several transition metal dichalcogenides (TMDCs).\cite{Miro2014,Butler2013}
There is a great variety of TMDCs, as many metal and chalcogen atoms can be combined to produce materials with properties that include metallic, 
semiconducting, and even superconducting behavior; the natural diversity of these materials with different properties make them  
particularly promising for electronic and optical applications.\cite{Wang2012a,MoS2book}
Unlike graphene, TMDCs such as MoS$_2$ and WS$_2$ have a finite band gap in the visible frequency range, which is indirect when in bulk (many layer) form, 
but becomes direct in the single 2D (trilayer) limit--where two S-layers are separated by a layer of Mo or W metal atoms.\cite{Mak2010a,Splendiani2010}
The direct gap in many of these single trilayer TMDCs makes them especially attractive candidates for optoelectronic and electronic applications,\cite{Britnell2013,Pospischil2014,Ross2014,Baugher2014}
such as field-effect transistors,\cite{Radisavljevic2011,Lembke2012,Wang2012} or photoaddressable sensors.\cite{Wang2012a}

Although we know a great deal about the electronic states in single trilayers, it is important to gain a detailed understanding of the electronic structure of finite size 
systems such as nanoribbons and quantum dots, in order to fully and reliably tailor the properties of different TMDC materials and possible devices.
Several works\cite{Bollinger2001a,Pan2012a,Erdogan:2012aa,Pavlovic2015b,Davelou2014a,Peterfalvi2015}
have reported the existence of  edge states in the gap of finite MoS$_2$ systems under different conditions.  The presence of metallic (dispersive) edge states in TMDCs nanostructures is especially relevant as new device geometries and interfaces become available; 
they would be expected to strongly affect transport  and optical properties of nanoribbons and 2D interfaces. \cite{Gong:2014aa,Chen:2015aa}

Edge or surface states also emerge in topological insulators, as has been intensely discussed in recent literature.\cite{Hasan2010,APLMat}
In those systems, it has been well established that the presence of edge states is a direct consequence of the principle of 
bulk-edge correspondence:\cite{Hasan2010,Teo2010} gapless states must be present at the domain 
wall separating two regions with different topological invariants.
Although pristine graphene is not a topological insulator due to its weak intrinsic spin-orbit interaction, the origin and
character of edge states in gapped and bilayer graphene 
has been analyzed in terms of the topological properties of the Hamiltonians for individual valleys.\cite{Li2011,Yao2009}
This analysis is made possible by the close analogy between graphene systems and 2D topological insulators.
The details of this analogy and its limitations have been discussed in the literature, but allow one to understand the appearance and
characteristics of symmetry allowed states at the edges of finite size systems. \cite{Li2010}
In light of the similar hexagonal structure of graphene and TMDCs, one may wonder if edge states in TMDCs single trilayers could 
be also analyzed in terms of the topological character imparted by the structure.

In this work, we use a two-band effective $\mathbf{k}\cdot \mathbf{p}$ model to investigate the electronic properties of 
MoS$_2$ nanoribbons and small triangular crystallites (`quantum dots') as those appearing naturally in growth chambers. 
We find the generic appearance of midgap states with wave functions strongly
localized near the edges of the structure, which can be clearly identified as edge states.  
Calculations for various sets of model parameters show that the appearance and characteristics of edge states are controlled by the curvature of
the 2D `bulk' band structure.
In particular, the sign of the band curvature parameters near the edge of the valence/conduction bands is found to be responsible for whether the 
edge states exist or not, and the relative magnitude of the effective masses determines the location of the states in the gap.
As in graphene systems,\cite{Li2010} all of these results can be understood as arising from the marginal topological properties of the MoS$_2$ 
single-valley Hamiltonians. In particular, we demonstrate that the Chern number per inequivalent valley is non vanishing in this structure, which
suggests the system may sustain edge states (and yet the system is topological trivial, with overall vanishing Chern number). 

The remainder of the paper is organized as follows. Sec.~\ref{sec:theory} presents the Hamiltonian used to describe the MoS$_2$ trilayers.
Then, in Sec.~\ref{sec:results}, we show and discuss typical numerical results for the two different systems under study:
MoS$_2$ nanoribbons (Sec.~\ref{sec:ribbon}) and MoS$_2$ quantum dots (Sec.~\ref{sec:QDs}), as defined by triangular crystallites.
Finally, conclusions are given in Sec.~\ref{sec:conclusions}.

\section{Theoretical model} \label{sec:theory}

As mentioned above, single trilayers of TMDC materials such as MoS$_2$ are composed by a layer of Mo atoms sandwiched between two layers of S atoms.
The metal atoms in this structure present trigonal prismatic coordination with the S atoms.  The electronic structure of the single trilayer has a direct 
gap at two non-equivalent points K and K' of the Brillouin zone.
Several works have derived an effective $\mathbf{k}\cdot \mathbf{p}$ model in the vicinity of these points
in order to study the low-energy physics of TMDC monolayers.\cite{Liu2013a,Kormanyos2013a,Rostami2013a,Kormanyos2015}
The proposed two-band Hamiltonian describing the valence and conduction bands up to second order in $k$ can be written as
\begin{equation}
\label{eq:ham}
H=
 \begin{pmatrix}
 \varepsilon_v + \alpha k^2 & \tau \gamma k_- \\
 \tau \gamma k_+ & \varepsilon_c + \beta k^2 \\
 \end{pmatrix},
\end{equation}

\noindent where $k_{\pm}=k_x \pm i \tau k_y$, and $\varepsilon_c=\Delta/2$ and $\varepsilon_v=-\Delta/2$ 
are the band-edge energies with $\Delta=1.9\,\mathrm{eV}$ standing for the material band gap;
$\mathbf{k}$ is the momentum relative to the K/K' points.  The constants
$\alpha$, $\beta$ and $\gamma$ are material parameters, while $\tau$ identifies the valley K ($\tau=1$) or K' ($\tau=-1$).

For the sake of simplicity, trigonal warping and other minor modifications present in the original model are neglected, although their inclusion would not
qualitatively alter the main conclusions of the work presented here.  
Hamiltonian~(\ref{eq:ham}) takes into account the electron-hole symmetry breaking obtained from first principles calculations by using 
unlike parameters $\alpha$ and $\beta$.
Although different authors report different values of these parameters, some dependent on the details of the calculations, we employ here
$\alpha=1.72\,\mathrm{eV \AA^2}$, $\beta=-0.13\,\mathrm{eV \AA^2}$ and $\gamma=3.82\,\mathrm{eV \AA}$, as
fitted from DFT calculations, \cite{Kormanyos2013a} unless noted otherwise.

Notice that (\ref{eq:ham}) ignores the spin degree of freedom, for clarity of presentation.  Consideration of spin-orbit coupling in these materials
results in effectively producing two valence band edges, as a spin-dependent gap appears, with corresponding spin-valley coupling in the valence band.  
The conduction band in MoS$_2$ has a 
sizable but relatively weaker spin-orbit splitting. \cite{Kosmider:2013qf,Kormanyos:2014fk}  Spin-orbit interactions will then result in a doubling
of the states we discuss here.  We revisit this issue in the discussion section below.  We also notice that 
the edges of the nanostructures are defined by hard-wall boundary conditions in all simulations, and are assumed to result in no 
intervalley coupling--as expected of zigzag edges.\cite{Peterfalvi2015}

\section{Numerical results and discussion} \label{sec:results}

We study the electronic properties of two different types of 2D nanostructures: \emph{nanoribbons}, where particles are
confined in one direction, and \emph{quantum dots}, where they are confined to triangular nanocrystallites.  The calculations are carried out
using COMSOL utilities over a fine grid (the finest default), and converged until the desired accuracy (typically 10$^{-12}$ in the eigenvalues).

\subsection{Nanoribbons}\label{sec:ribbon}

\begin{figure}[h!]
\begin{center}
\includegraphics[width=0.45\textwidth]{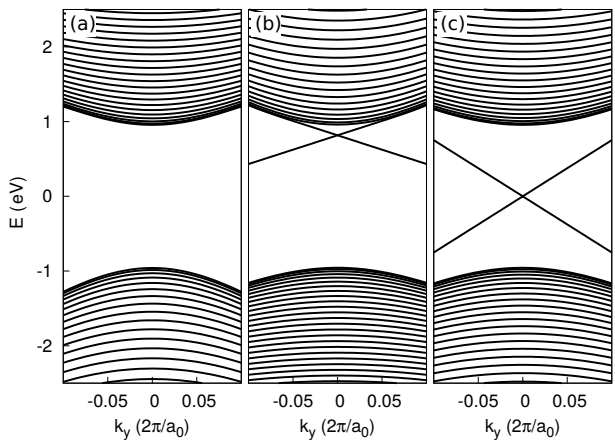}
\caption{Energy band dispersion for MoS$_2$ nanoribbons considering different values of $\alpha$ and $\beta$: 
(a) $\alpha=-1.72\,\mathrm{eV \AA^2}$ and $\beta=0.13\,\mathrm{eV \AA^2}$,
(b) $\alpha=1.72\,\mathrm{eV \AA^2}$ and $\beta=-0.13\,\mathrm{eV \AA^2}$, and
(c) $\alpha=1.72\,\mathrm{eV \AA^2}$ and $\beta=-1.72\,\mathrm{eV \AA^2}$.
The edges are parallel to the $y$ direction, and the wave vector $k_y$ is measured with respect to the K valley, where $a_0=3.193\,\mathrm{\AA}$ is the lattice constant.} 
\label{fig1}
\end{center}
\end{figure}

The nanoribbons are defined over a finite width along the direction $x$ in our calculations, while maintaining translational invariance along the $y$ direction. 
As such, the momentum $k_y$ is a good quantum number and the two-component spinor wave function of Hamiltonian~(\ref{eq:ham}) can be written in the form
$\psi (x,y)=e^{i k_y y} \phi(x)$, where $\psi$ and $\phi$ have components over the $c,v$ basis.
As a consequence, the eigenvalue equation of this 2D Hamiltonian turns into a set of two coupled second-order differential equations in 
one-dimension that depend on the quantum number $k_y$.
We solve numerically these equations for an MoS$_2$ nanoribbon of 10 nm width, wide enough to allow decoupled states on both edges, as we will see. 
The results obtained are summarized in Fig.~\ref{fig1}.

Fig.~\ref{fig1}(b) shows the calculated subband dispersion.  Notice that the finite width of the ribbon has only slightly opened the gap, as the effective masses
near the band edges, $m_v$ and $m_c$, are both $\approx 0.5$, and the size quantization is only a few meV.\@  
Most importantly, we find two states inside the band gap, with a nearly linear dispersion.
The levels cross at $k_y=0$ and $E=0.816\,\mathrm{eV}$, relatively close to the edge of the conduction band. 
These midgap states disperse upwards in energy, close to the conduction band for not large
$k_y$ values ($k_y \approx  \pm 0.05(2\pi/a_0)$, see Fig.~\ref{fig1}b), and soon admix with the band states,  becoming indistinguishable from them.
For lower energies, however, the midgap states remain well defined and exhibit increasing edge localization, as we will see below. 

In order to study the origin of these states, and dependence on band structure features,
we carry out the same calculations but for other sets of parameters than those in
Ref.~\onlinecite{Kormanyos2013a}.
We only tune the $\alpha$ and $\beta$ values since $\gamma$ does not qualitatively affect the results.
In Fig.~\ref{fig1}(a) we exchange the signs of both $\alpha$ and $\beta$ 
($\alpha=-1.72\,\mathrm{eV \AA^2}$ and $\beta=0.13\,\mathrm{eV \AA^2}$) 
and observe that the states lying inside the gap disappear.
In Fig.~\ref{fig1}(c) we keep the signs unaltered to those in panel (b) but modify $\beta$ to have the same absolute value of $\alpha$
 ($\alpha=1.72\,\mathrm{eV \AA^2}$ and $\beta=-1.72\,\mathrm{eV \AA^2}$).
In this case, the two states inside the gap are still present but they have lower energies compared to Fig.~\ref{fig1}(b).
As expected from symmetry, the dispersion bands now cross at $k_y=0$ and $E=0$, 
since $\alpha=\beta$ confers electron-hole symmetry to the Hamiltonian.

By comparing the results in Fig.~\ref{fig1} for the three sets of parameters, it is clear that the presence or absence of 
states inside the gap is determined by the sign of both $\alpha$ and $\beta$ curvatures.
Midgap states exist if $\alpha>0$ and $\beta<0$ and are absent if $\alpha<0$ and $\beta>0$.\footnote{We do not consider 
parameters $\alpha$ and $\beta$ with the same sign since in these cases there is not a real gap separating the bands.}
Furthermore, changes in the relative value of these two parameters affect the energy of the states inside the gap.
When $|\alpha|>|\beta|$ the states are closer to the conduction band as in Fig.~\ref{fig1}(b), and when $|\alpha|<|\beta|$
they become closer to the valence band.

One can qualitatively analyze this behavior in terms of the `bare' effective masses for valence and conduction bands, as determined
by the $\alpha$ and $\beta$ coefficients.  A negative $\beta$ (and corresponding negative mass $\simeq 1/\beta$) in the conduction band
is `inverted', and that symmetry is contained in the states even after the mixing due to $\gamma$.  The inverse effective masses for the full Hamiltonian (\ref{eq:ham})
near the edges are however given by $2(\beta + \gamma^2/\Delta)/\hbar^2$ for the
conduction band, and by $2(\alpha - \gamma^2/\Delta)/\hbar^2$ for the valence band, and therefore dominated by the large value of $\gamma$. 

\begin{figure}[h!]
\begin{center}
\includegraphics[width=0.5\textwidth]{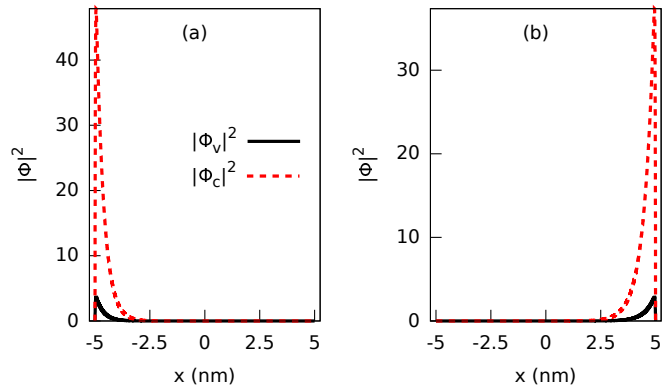}
\caption{Wave function squared modulus $|\phi|^2$ of the two states at $k_y= 0.01 \times 2\pi/a_0$, with energies lying in the band gap (a) $E=0.778\,\mathrm{eV}$, and (b) $E=0.855\,\mathrm{eV}$.
Black solid lines correspond to the valence band component and red dashed lines to the conduction band.}
\label{fig2}
\end{center}
\end{figure}

To further explore the nature of the states inside the gap, we analyze the wave functions in Fig.~\ref{fig2}.
As an example, we choose the states for $k_y= 0.01 \times 2\pi/a_0$ in Fig.~\ref{fig1}(b), which are slightly 
away from the degeneracy point, and well away from the conduction band states.
Fig.~\ref{fig2}(a) corresponds to the lower state at $E=0.778\,\mathrm{eV}$ and Fig.~\ref{fig2}(b) to the higher one at $E=0.855\,\mathrm{eV}$.
We clearly observe that both states are localized at opposite edges of the MoS$_2$ nanoribbon--and have
opposite dispersion, as expected of independent edge states.
We see that the conduction band component (red dashed line) is the dominant contribution to the wave function.
Calculation of the relative weight of the two components yields $w(\phi_c)=93\%$ and $w(\phi_v)=7\%$ for the conduction and valence band components, respectively.
These values can be directly obtained from the parameters $\alpha$ and $\beta$ using the following expressions $w(\phi_c)=|\beta|/(|\alpha|+|\beta|)$ and 
$w(\phi_v)=|\alpha|/(|\alpha|+|\beta|)$.  
These expressions hold as long as the edge states are relatively far from the bulk bands.
Notice also the asymmetry in the wave functions as seen, for instance, in the different maximum 
value of $|\phi_c|^2$, and their different $x$-extension.  
This asymmetry is due to the proximity of the conduction band.
The higher energy edge state is slightly more admixed with the bulk states and, thus, 
its wave function results somewhat more delocalized.  
The asymmetry in the states continues to grow as $k_y$ increases further.

The results summarized in Fig.~\ref{fig1} and Fig.~\ref{fig2} can be related to those coming from the 
model proposed by Bernevig, Hughes and Zhang (BHZ),\cite{BHZ2006}
in connection with the observation of the quantum spin Hall effect (QSHE).
In that work, the QSHE was predicted in HgTe quantum wells larger than a critical thickness,
due to a band inversion in the low energy effective Hamiltonian.
For $\Delta<0$, bands are inverted and the system shows topological behavior.
One consequence is that edge states will form when a transition between two distinct topological phases takes place, 
as predicted by the principle of bulk-edge correspondence.\cite{Hasan2010}
In our system, Eq.\ (\ref{eq:ham}), we have $\Delta>0$, which is apparently trivial, 
although the sign of the bare band curvatures ($\alpha>0$ and $\beta<0$) yields
also a situation with inverted bands.  As such, the origin of the edge states here can be analyzed in terms of 
the topological character of the model in (\ref{eq:ham}).

\begin{figure}[h!]
\begin{center}
\includegraphics[width=0.45\textwidth]{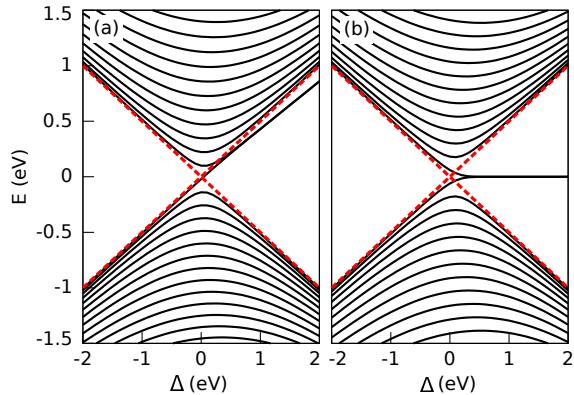}
\caption{Energy spectrum of a MoS$_2$ nanoribbon as in Fig.\ \ref{fig1}, 
shown as a function of the bandgap $\Delta$, for $k_y=0$.
Two sets of parameters are considered: (a)  $\alpha=1.72\,\mathrm{eV \AA^2}$ and $\beta=-0.13\,\mathrm{eV \AA^2}$, and
(b) $\alpha=1.72\,\mathrm{eV \AA^2}=-\beta$.
Red dashed lines indicate the edges of the bandgap.  Midgap edge states appear for $\Delta >0$ in both cases.}
\label{fig3}
\end{center}
\end{figure}

To explore this relationship further, Fig.~\ref{fig3} shows the energy spectrum as a function of $\Delta$, for 
a given set of $\alpha$ and $\beta$ parameters.
The spectra shown are for $k_y=0$ and band curvatures $\alpha=1.72\,\mathrm{eV \AA^2}$ and $\beta=-0.13\,\mathrm{eV \AA^2}$ in Fig.~\ref{fig3}(a),
and $\alpha=1.72\,\mathrm{eV \AA^2}$ and $\beta=-1.72\,\mathrm{eV \AA^2}$ in \ref{fig3}(b).
Two red dashed lines in each panel show the limits of the band gap, for reference.
In both cases shown, we see that a trivial situation develops, with no states in the gap, for negative $\Delta$ values. 
As $\Delta$ increases and changes to positive values, the conduction and valence bands 
seem to be similar, except for the appearance of a pair (for $k_y=0$) of 
degenerate edge states with energies clearly in the gap.  Notice that the edge states separate from the
conduction band for larger $\Delta$ values in Fig.\ \ref{fig3}(a), but remain exactly equidistant from 
both bands for $|\alpha|=|\beta|$ in \ref{fig3}(b), as expected, appearing closer to 
the conduction band for more asymmetric $|\alpha|>|\beta|$ values.

Next, we look at these results with the help of the Chern number associated with the occupied band 
(topological invariant).
For a two-level Hamiltonian written in the form $H(\mathbf{k})=\mathbf{g}(\mathbf{k}) \cdot \boldsymbol{\sigma}$, where $\boldsymbol{\sigma}$ is a vector with the 
Pauli matrices as components, the Chern number is given by\cite{Hasan2010}
\begin{equation}
c=\frac{1}{4 \pi} \int d^2k ( \partial_{k_x} \mathbf{\hat{g}} \times \partial_{k_y} \mathbf{\hat{g}} ) \cdot \mathbf{\hat{g}} \, ,
\end{equation}

\noindent where $\mathbf{\hat{g}}=\mathbf{g}/|\mathbf{g}|$ and the integral is computed over the entire Brillouin zone.
For the Hamiltonian in Eq.~(\ref{eq:ham}), one obtains $c=\tau/2 \big(\sgn{(\Delta)} + \sgn{(\alpha-\beta)}\big)$, fully independent of the value of $\gamma$. 
That means that for $\Delta>0$, one obtains $c=0$
for $\alpha< \beta$, while $c=\tau$ for $\alpha>\beta$.  A non zero value of $c$ suggests that
a topological invariant is present, and this goes along with the previous discussion based on 
band inversion arguments.
It is important to note, however, that the contribution of the K and K' valleys to the topological
invariant has opposite signs, which produces an overall $c=0$.
As a result, one can strictly state that multivalley materials such as graphene or MoS$_2$ are topologically trivial.

In spite of the strict trivial topology of Eq.\ ({\ref{eq:ham}), a non-vanishing $c$ for a single-valley  
can be associated with marginal topological properties, in analogy with topological insulators.\cite{Li2010}
This analogy has however important limitations.
Since $c$ per valley is not a well-defined topological invariant, $c \neq 0$ does not guarantee the 
existence of edge states at the boundaries with the vacuum.
Furthermore, and perhaps most important, is the fact that if edge states are present, they are not topologically protected against backscattering and can then be affected by
any type of disorder and/or valley coupling.  Nevertheless, it is the case that edge states in bilayer graphene 
have been shown to be robust under moderate disorder,\cite{Li2011} and to exhibit pure valley currents, as
indicated by the local valley Berry curvature and associated Chern number. \cite{Gorbachev2014,Shimazaki2015}

We should also comment that although, for simplicity, we have suppressed the spin degree of freedom in these calculations, its role can
be easily established.  The presence of spin clearly results in \emph{two} edge states per border of the structure, instead
of the single state presented above--see Fig.\ \ref{fig1}(b).  As the spin-orbit interaction in the valence band is large (yet much smaller than the bandgap, 
and diagonal, pinning the spin projection to each of the valleys), the two edge states on the same border but different spin projection appear 
slightly shifted in energy and minimally different dispersion (not shown).  This
simple duplication of edge states with different spins and energies would of course be strongly affected if the edges couple valleys,
something that will depend on the border terminations and corresponding boundary conditions. \cite{Erdogan:2012aa,Peterfalvi2015}

\subsection{MoS$_2$ triangular quantum dots} \label{sec:QDs}

We next investigate the electronic properties of MoS$_2$ quantum dots formed by finite size flakes.
The flakes are equilateral triangles, as it is a commonly synthesized shape.\cite{Duan2014,Marshall2014,Liu2014,Bao2015}
In this case, carriers are confined in the two directions of space and we must numerically solve the 
coupled differential equations in two variables
in order to find eigenvalues and eigenfunctions of the Hamiltonian. In the results that follow, Fig.\ \ref{fig4},
the quantum dot side length is $10\,\mathrm{nm}$, and we 
employ the same parameters as in the previous section--see Fig.\ \ref{fig1}(b).

\begin{figure}[!h]
\begin{center}
\includegraphics[width=0.40\textwidth]{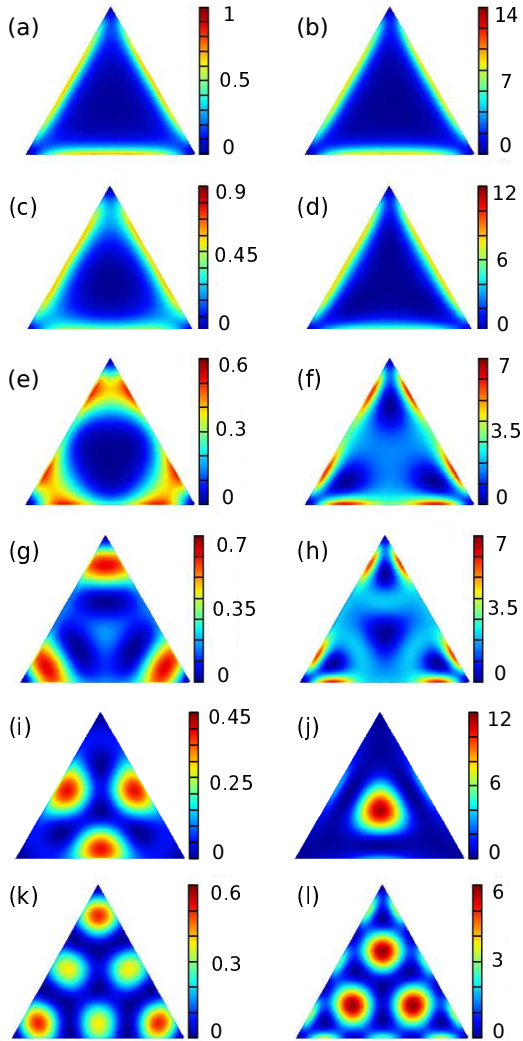}
\caption{Wave function squared modulus of selected states with energies close to the conduction band.
Left and right columns illustrate the valence $|\phi_v|^2$ and conduction band $|\phi_c|^2$ components, 
respectively. Different states are arranged in rows with increasing energy: 
(a)-(b) $E=0.907\,\mathrm{eV}$, (c)-(d) $E=0.962\,\mathrm{eV}$, (e)-(f) $E=1.015\,\mathrm{eV}$,
(g)-(h) $E=1.022\,\mathrm{eV}$, (i)-(j) $E=1.057\,\mathrm{eV}$ and (k)-(l) $E=1.100\,\mathrm{eV}$.
As before, $|\phi_c|^2$ components are generally larger for these states, close to the conduction band.
}
\label{fig4}
\end{center}
\end{figure}

The results obtained for this system show the presence of several states with energies in the gap. 
They can be seen as the result of the 
discretization of the edge states along each border, which are then hybridized near the corners of the flake. 
We illustrate this behavior in Fig.~\ref{fig4}, where the squared modulus of the wave function for a selection 
of states with energy close to the conduction band ($E \approx 0.95\,\mathrm{eV}$) is shown.
We choose this range of energies because we know that for $\alpha=1.72\,\mathrm{eV \AA^2}$ and $\beta=-0.13\,\mathrm{eV \AA^2}$, \cite{Kormanyos2013a}
the edge states are closer to the conduction band.
By gradually increasing energy, we also get to compare the clearly `bulk' and edge states in the flake.
In Fig.~\ref{fig4}(a)-(d) we can see that the first two states are clearly edge states with wave functions 
localized near the triangle border, with similar appearance to that shown in Fig.\ \ref{fig2} near each of
the edges. 
The next two states in energy, Fig.~\ref{fig4}(e)-(h), also show wave functions mainly near the edges,
but noticeably more delocalized than the previous two.
This suggests that the edge states are partially admixed with the bulk conduction band, due to their energy proximity.
Finally, Fig.~\ref{fig4}(i)-(l) shows two conduction states with wave functions completely delocalized over the entire
triangular quantum dot.
A representation of the real and imaginary parts of the wave functions (not shown) allows one to see the wave function nodes more easily, and see that their number increases with the energy of the states, an expected signature of quantization.

We should emphasize that we have also found the same pattern of
edge states appearing for the curvature parameters as in the case of nanoribbons.  As
such, one can also invoke the marginal topological character
of the Hamiltonian as the origin of edge states in these zero-dimensional nanostructrures.

\section{Conclusions} \label{sec:conclusions}

The low-energy electronic structure of monolayer MoS$_2$ nanoribbons and quantum dots has been investigated using an effective
two-band $\mathbf{k}\cdot \mathbf{p}$ model.
We have shown that both systems present edge states with energies in the gap.
Nanoribbons exhibit only one state per edge at a given value of the quantum number $k_y$, while in quantum dots, 
due to full confinement, the edge states appear distinctly away from the states that would form the subband continuum in a
large triangular flake.  As the energy of the edge states increases, for both nanoribbons and quantum dots, the edge states
hybridize with the `bulk' and cease to be so well-localized near the edges of the structure.   

We have also found that the curvature of the bands, represented by parameters $\alpha$ and $\beta$, 
determine the presence ($\alpha>0$ and $\beta<0$) or absence ($\alpha<0$ and $\beta>0$)
of edge states as well as their energy.  This behavior is reminiscent of the marginal topological properties of materials such as MoS$_2$,
as the Chern number per valley is indeed non-zero, reflecting a finite Berry curvature in each valley.  
Similar results are of course expected for other TMDC nanostructures, as long as the relative band curvature parameter signs are similar to
those presented here.

We should emphasize that first principles calculations are typically used to determine continuum model parameters.  As the former may depend
on functionals and other details of the calculations, the latter may indeed vary among different implementations and/or authors.  In fact, some
fittings result in values of $\alpha$ and $\beta$ that are indeed substantially different, and for which the edge states we discuss here are not
apparent. \cite{Kormanyos2013a}  It is also clear that tight-binding parameterizations may similarly allow for the presence of edge states, as 
explicitly seen in the literature. \cite{Pavlovic2015b}  These edge states, however, are found to be rather robust and to exist over a wide
range of parameters.

Considering their interesting characteristics, it would be interesting to explore experimentally how truly robust these edge states are in 
nanoribbons and other natural structures with edges.
Exploring what observable consequences they have on the effective trapping of photo-activated carriers and excitons, or
how they modulate interactions between adsorbed/embedded impurity atoms, may provide further insights into the appearance of edge states in these systems.

\section*{Acknowledgements}
This work was supported by Spanish FPU Grant (CS).  Support from  NSF grant DMR-1508325 (SEU), 
UJI Project No.\ P1.1B2014-24 and MINECO Project No.\ CTQ2014-60178-P (JP) is acknowledged.


\begin{thebibliography}{10}%
\makeatletter
\providecommand \@ifxundefined [1]{%
 \ifx #1\undefined \expandafter \@firstoftwo
 \else \expandafter \@secondoftwo
\fi
}%
\providecommand \@ifnum [1]{%
 \ifnum #1\expandafter \@firstoftwo
 \else \expandafter \@secondoftwo
\fi
}%
\providecommand \enquote [1]{``#1''}%
\providecommand \bibnamefont  [1]{#1}%
\providecommand \bibfnamefont [1]{#1}%
\providecommand \citenamefont [1]{#1}%
\providecommand\href[0]{\@sanitize\@href}%
\providecommand\@href[1]{\endgroup\@@startlink{#1}\endgroup\@@href}%
\providecommand\@@href[1]{#1\@@endlink}%
\providecommand \@sanitize [0]{\begingroup\catcode`\&12\catcode`\#12\relax}%
\@ifxundefined \pdfoutput {\@firstoftwo}{%
 \@ifnum{\z@=\pdfoutput}{\@firstoftwo}{\@secondoftwo}%
}{%
 \providecommand\@@startlink[1]{\leavevmode}%
 \providecommand\@@endlink[0]{}%
}{%
 \providecommand\@@startlink[1]{%
  \leavevmode
  \pdfstartlink
   attr{/Border[0 0 1 ]/H/I/C[0 1 1]}%
   user{/Subtype/Link/A<</Type/Action/S/URI/URI(#1)>>}%
  \relax
 }%
 \providecommand\@@endlink[0]{\pdfendlink}%
}%
\providecommand \url  [0]{\begingroup\@sanitize \@url }%
\providecommand \@url [1]{\endgroup\@href {#1}{\urlprefix}}%
\providecommand \urlprefix [0]{URL }%
\providecommand \Eprint[0]{\href }%
\@ifxundefined \urlstyle {%
  \providecommand \doi [1]{doi:\discretionary{}{}{}#1}%
}{%
  \providecommand \doi [0]{doi:\discretionary{}{}{}\begingroup
  \urlstyle{rm}\Url }%
}%
\providecommand \doibase [0]{http://dx.doi.org/}%
\providecommand \Doi[1]{\href{\doibase#1}}%
\providecommand \bibAnnote [3]{%
  \BibitemShut{#1}%
  \begin{quotation}\noindent
    \textsc{Key:}\ #2\\\textsc{Annotation:}\ #3%
  \end{quotation}%
}%
\providecommand \bibAnnoteFile [2]{%
  \IfFileExists{#2}{\bibAnnote {#1} {#2} {\input{#2}}}{}%
}%
\providecommand \typeout [0]{\immediate \write \m@ne }%
\providecommand \selectlanguage [0]{\@gobble}%
\providecommand \bibinfo [0]{\@secondoftwo}%
\providecommand \bibfield [0]{\@secondoftwo}%
\providecommand \translation [1]{[#1]}%
\providecommand \BibitemOpen[0]{}%
\providecommand \bibitemStop [0]{}%
\providecommand \bibitemNoStop [0]{.\EOS\space}%
\providecommand \EOS [0]{\spacefactor3000\relax}%
\providecommand \BibitemShut [1]{\csname bibitem#1\endcsname}%
\bibitem{Novoselov2004}%
  \BibitemOpen
  \bibfield{author}{%
  \bibinfo {author} {\bibfnamefont{K.~S.}\ \bibnamefont{Novoselov}}, \bibinfo
  {author} {\bibfnamefont{A.~K.}\ \bibnamefont{Geim}}, \bibinfo {author}
  {\bibfnamefont{S.~V.}\ \bibnamefont{Morozov}}, \bibinfo {author}
  {\bibfnamefont{D.}~\bibnamefont{Jiang}}, \bibinfo {author}
  {\bibfnamefont{Y.}~\bibnamefont{Zhang}}, \bibinfo {author}
  {\bibfnamefont{S.~V.}\ \bibnamefont{Dubonos}}, \bibinfo {author}
  {\bibfnamefont{I.~V.}\ \bibnamefont{Grigorieva}},\ and\ \bibinfo {author}
  {\bibfnamefont{A.~A.}\ \bibnamefont{Firsov}},\ }%
  \bibfield{journal}{%
  \bibinfo {journal} {Science}\ }%
  \textbf{\bibinfo {volume} {306}},\ \bibinfo {pages} {666} (\bibinfo {year}
  {2004})%
  \bibAnnoteFile{NoStop}{Novoselov2004}%
\bibitem{Novoselov:2005aa}%
  \BibitemOpen
  \bibfield{author}{%
  \bibinfo {author} {\bibfnamefont{K.~S.}\ \bibnamefont{Novoselov}}, \bibinfo
  {author} {\bibfnamefont{D.}~\bibnamefont{Jiang}}, \bibinfo {author}
  {\bibfnamefont{F.}~\bibnamefont{Schedin}}, \bibinfo {author}
  {\bibfnamefont{T.~J.}\ \bibnamefont{Booth}}, \bibinfo {author}
  {\bibfnamefont{V.~V.}\ \bibnamefont{Khotkevich}}, \bibinfo {author}
  {\bibfnamefont{S.~V.}\ \bibnamefont{Morozov}},\ and\ \bibinfo {author}
  {\bibfnamefont{A.~K.}\ \bibnamefont{Geim}},\ }%
  \bibfield{journal}{%
  \bibinfo {journal} {Proc. Natl. Acad. U.S.A.}\ }%
  \textbf{\bibinfo {volume} {102}},\ \bibinfo {pages} {10451} (\bibinfo {month}
  {07}\ \bibinfo {year} {2005})%
  \bibAnnoteFile{NoStop}{Novoselov:2005aa}%
\bibitem{Miro2014}%
  \BibitemOpen
  \bibfield{author}{%
  \bibinfo {author} {\bibfnamefont{P.}~\bibnamefont{Mir\'{o}}}, \bibinfo
  {author} {\bibfnamefont{M.}~\bibnamefont{Audiffred}},\ and\ \bibinfo {author}
  {\bibfnamefont{T.}~\bibnamefont{Heine}},\ }%
  \bibfield{journal}{%
  \bibinfo {journal} {Chem. Soc. Rev.}\ }%
  \textbf{\bibinfo {volume} {43}},\ \bibinfo {pages} {6537} (\bibinfo {year}
  {2014})%
  \bibAnnoteFile{NoStop}{Miro2014}%
\bibitem{Butler2013}%
  \BibitemOpen
  \bibfield{author}{%
  \bibinfo {author} {\bibfnamefont{S.~Z.}\ \bibnamefont{Butler}}, \bibinfo
  {author} {\bibfnamefont{S.~M.}\ \bibnamefont{Hollen}}, \bibinfo {author}
  {\bibfnamefont{L.}~\bibnamefont{Cao}}, \bibinfo {author}
  {\bibfnamefont{Y.}~\bibnamefont{Cui}}, \bibinfo {author}
  {\bibfnamefont{J.~A.}\ \bibnamefont{Gupta}}, \bibinfo {author}
  {\bibfnamefont{H.~R.}\ \bibnamefont{Guti\'{e}rrez}}, \bibinfo {author}
  {\bibfnamefont{T.~F.}\ \bibnamefont{Heinz}}, \bibinfo {author}
  {\bibfnamefont{S.~S.}\ \bibnamefont{Hong}}, \bibinfo {author}
  {\bibfnamefont{J.}~\bibnamefont{Huang}}, \bibinfo {author}
  {\bibfnamefont{A.~F.}\ \bibnamefont{Ismach}}, \bibinfo {author}
  {\bibfnamefont{E.}~\bibnamefont{Johnston-Halperin}}, \bibinfo {author}
  {\bibfnamefont{M.}~\bibnamefont{Kuno}}, \bibinfo {author}
  {\bibfnamefont{V.~V.}\ \bibnamefont{Plashnitsa}}, \bibinfo {author}
  {\bibfnamefont{R.~D.}\ \bibnamefont{Robinson}}, \bibinfo {author}
  {\bibfnamefont{R.~S.}\ \bibnamefont{Ruoff}}, \bibinfo {author}
  {\bibfnamefont{S.}~\bibnamefont{Salahuddin}}, \bibinfo {author}
  {\bibfnamefont{J.}~\bibnamefont{Shan}}, \bibinfo {author}
  {\bibfnamefont{L.}~\bibnamefont{Shi}}, \bibinfo {author}
  {\bibfnamefont{M.~G.}\ \bibnamefont{Spencer}}, \bibinfo {author}
  {\bibfnamefont{M.}~\bibnamefont{Terrones}}, \bibinfo {author}
  {\bibfnamefont{W.}~\bibnamefont{Windl}},\ and\ \bibinfo {author}
  {\bibfnamefont{J.~E.}\ \bibnamefont{Goldberger}},\ }%
  \bibfield{journal}{%
  \bibinfo {journal} {ACS Nano}\ }%
  \textbf{\bibinfo {volume} {7}},\ \bibinfo {pages} {2898} (\bibinfo {year}
  {2013})%
  \bibAnnoteFile{NoStop}{Butler2013}%
\bibitem{Wang2012a}%
  \BibitemOpen
  \bibfield{author}{%
  \bibinfo {author} {\bibfnamefont{Q.~H.}\ \bibnamefont{Wang}}, \bibinfo
  {author} {\bibfnamefont{K.}~\bibnamefont{Kalantar-Zadeh}}, \bibinfo {author}
  {\bibfnamefont{A.}~\bibnamefont{Kis}}, \bibinfo {author}
  {\bibfnamefont{J.~N.}\ \bibnamefont{Coleman}},\ and\ \bibinfo {author}
  {\bibfnamefont{M.~S.}\ \bibnamefont{Strano}},\ }%
  \bibfield{journal}{%
  \bibinfo {journal} {Nat. Nanotechnol.}\ }%
  \textbf{\bibinfo {volume} {7}},\ \bibinfo {pages} {699} (\bibinfo {year}
  {2012})%
  \bibAnnoteFile{NoStop}{Wang2012a}%
\bibitem{MoS2book}%
  \BibitemOpen
  \bibfield{author}{%
  \bibinfo {author} {\bibfnamefont{Z.~M.}\ \bibnamefont{Wang}},\ }%
  \emph{\bibinfo {title} {{MoS$_2$: Materials, Physics, and Devices}}}\
  (\bibinfo {publisher} {Springer International},\ \bibinfo {address}
  {Switzerland},\ \bibinfo {year} {2014})%
  \bibAnnoteFile{NoStop}{MoS2book}%
\bibitem{Mak2010a}%
  \BibitemOpen
  \bibfield{author}{%
  \bibinfo {author} {\bibfnamefont{K.~F.}\ \bibnamefont{Mak}}, \bibinfo
  {author} {\bibfnamefont{C.}~\bibnamefont{Lee}}, \bibinfo {author}
  {\bibfnamefont{J.}~\bibnamefont{Hone}}, \bibinfo {author}
  {\bibfnamefont{J.}~\bibnamefont{Shan}},\ and\ \bibinfo {author}
  {\bibfnamefont{T.~F.}\ \bibnamefont{Heinz}},\ }%
  \bibfield{journal}{%
  \bibinfo {journal} {Phys. Rev. Lett.}\ }%
  \textbf{\bibinfo {volume} {105}},\ \bibinfo {pages} {136805} (\bibinfo {year}
  {2010})%
  \bibAnnoteFile{NoStop}{Mak2010a}%
\bibitem{Splendiani2010}%
  \BibitemOpen
  \bibfield{author}{%
  \bibinfo {author} {\bibfnamefont{A.}~\bibnamefont{Splendiani}}, \bibinfo
  {author} {\bibfnamefont{L.}~\bibnamefont{Sun}}, \bibinfo {author}
  {\bibfnamefont{Y.}~\bibnamefont{Zhang}}, \bibinfo {author}
  {\bibfnamefont{T.}~\bibnamefont{Li}}, \bibinfo {author}
  {\bibfnamefont{J.}~\bibnamefont{Kim}}, \bibinfo {author}
  {\bibfnamefont{C.~Y.}\ \bibnamefont{Chim}}, \bibinfo {author}
  {\bibfnamefont{G.}~\bibnamefont{Galli}},\ and\ \bibinfo {author}
  {\bibfnamefont{F.}~\bibnamefont{Wang}},\ }%
  \bibfield{journal}{%
  \bibinfo {journal} {Nano Lett.}\ }%
  \textbf{\bibinfo {volume} {10}},\ \bibinfo {pages} {1271} (\bibinfo {year}
  {2010})%
  \bibAnnoteFile{NoStop}{Splendiani2010}%
\bibitem{Britnell2013}%
  \BibitemOpen
  \bibfield{author}{%
  \bibinfo {author} {\bibfnamefont{L.}~\bibnamefont{Britnell}}, \bibinfo
  {author} {\bibfnamefont{R.~M.}\ \bibnamefont{Ribeiro}}, \bibinfo {author}
  {\bibfnamefont{A.}~\bibnamefont{Eckmann}}, \bibinfo {author}
  {\bibfnamefont{R.}~\bibnamefont{Jalil}}, \bibinfo {author}
  {\bibfnamefont{B.~D.}\ \bibnamefont{Belle}}, \bibinfo {author}
  {\bibfnamefont{A.}~\bibnamefont{Mishchenko}}, \bibinfo {author}
  {\bibfnamefont{Y.-J.}\ \bibnamefont{Kim}}, \bibinfo {author}
  {\bibfnamefont{R.~V.}\ \bibnamefont{Gorbachev}}, \bibinfo {author}
  {\bibfnamefont{T.}~\bibnamefont{Georgiou}}, \bibinfo {author}
  {\bibfnamefont{S.~V.}\ \bibnamefont{Morozov}}, \bibinfo {author}
  {\bibfnamefont{A.~N.}\ \bibnamefont{Grigorenko}}, \bibinfo {author}
  {\bibfnamefont{A.~K.}\ \bibnamefont{Geim}}, \bibinfo {author}
  {\bibfnamefont{C.}~\bibnamefont{Casiraghi}}, \bibinfo {author}
  {\bibfnamefont{A.~H.}\ \bibnamefont{Castro~Neto}},\ and\ \bibinfo {author}
  {\bibfnamefont{K.~S.}\ \bibnamefont{Novoselov}},\ }%
  \bibfield{journal}{%
  \bibinfo {journal} {Science}\ }%
  \textbf{\bibinfo {volume} {340}},\ \bibinfo {pages} {1311} (\bibinfo {year}
  {2013})%
  \bibAnnoteFile{NoStop}{Britnell2013}%
\bibitem{Pospischil2014}%
  \BibitemOpen
  \bibfield{author}{%
  \bibinfo {author} {\bibfnamefont{A.}~\bibnamefont{Pospischil}}, \bibinfo
  {author} {\bibfnamefont{M.~M.}\ \bibnamefont{Furchi}},\ and\ \bibinfo
  {author} {\bibfnamefont{T.}~\bibnamefont{Mueller}},\ }%
  \bibfield{journal}{%
  \bibinfo {journal} {Nat. Nanotechnol.}\ }%
  \textbf{\bibinfo {volume} {9}},\ \bibinfo {pages} {257} (\bibinfo {year}
  {2014})%
  \bibAnnoteFile{NoStop}{Pospischil2014}%
\bibitem{Ross2014}%
  \BibitemOpen
  \bibfield{author}{%
  \bibinfo {author} {\bibfnamefont{J.~S.}\ \bibnamefont{Ross}}, \bibinfo
  {author} {\bibfnamefont{P.}~\bibnamefont{Klement}}, \bibinfo {author}
  {\bibfnamefont{A.~M.}\ \bibnamefont{Jones}}, \bibinfo {author}
  {\bibfnamefont{N.~J.}\ \bibnamefont{Ghimire}}, \bibinfo {author}
  {\bibfnamefont{J.}~\bibnamefont{Yan}}, \bibinfo {author}
  {\bibfnamefont{D.~G.}\ \bibnamefont{Mandrus}}, \bibinfo {author}
  {\bibfnamefont{T.}~\bibnamefont{Taniguchi}}, \bibinfo {author}
  {\bibfnamefont{K.}~\bibnamefont{Watanabe}}, \bibinfo {author}
  {\bibfnamefont{K.}~\bibnamefont{Kitamura}}, \bibinfo {author}
  {\bibfnamefont{W.}~\bibnamefont{Yao}}, \bibinfo {author}
  {\bibfnamefont{D.~H.}\ \bibnamefont{Cobden}},\ and\ \bibinfo {author}
  {\bibfnamefont{X.}~\bibnamefont{Xu}},\ }%
  \bibfield{journal}{%
  \bibinfo {journal} {Nat. Nanotechnol.}\ }%
  \textbf{\bibinfo {volume} {9}},\ \bibinfo {pages} {268} (\bibinfo {year}
  {2014})%
  \bibAnnoteFile{NoStop}{Ross2014}%
\bibitem{Baugher2014}%
  \BibitemOpen
  \bibfield{author}{%
  \bibinfo {author} {\bibfnamefont{B.~W.~H.}\ \bibnamefont{Baugher}}, \bibinfo
  {author} {\bibfnamefont{H.~O.~H.}\ \bibnamefont{Churchill}}, \bibinfo
  {author} {\bibfnamefont{Y.}~\bibnamefont{Yang}},\ and\ \bibinfo {author}
  {\bibfnamefont{P.}~\bibnamefont{Jarillo-Herrero}},\ }%
  \bibfield{journal}{%
  \bibinfo {journal} {Nat. Nanotechnol.}\ }%
  \textbf{\bibinfo {volume} {9}},\ \bibinfo {pages} {262} (\bibinfo {year}
  {2014})%
  \bibAnnoteFile{NoStop}{Baugher2014}%
\bibitem{Radisavljevic2011}%
  \BibitemOpen
  \bibfield{author}{%
  \bibinfo {author} {\bibfnamefont{B.}~\bibnamefont{Radisavljevic}}, \bibinfo
  {author} {\bibfnamefont{A.}~\bibnamefont{Radenovic}}, \bibinfo {author}
  {\bibfnamefont{J.}~\bibnamefont{Brivio}}, \bibinfo {author}
  {\bibfnamefont{V.}~\bibnamefont{Giacometti}},\ and\ \bibinfo {author}
  {\bibfnamefont{A.}~\bibnamefont{Kis}},\ }%
  \bibfield{journal}{%
  \bibinfo {journal} {Nat. Nanotechnol.}\ }%
  \textbf{\bibinfo {volume} {6}},\ \bibinfo {pages} {147} (\bibinfo {year}
  {2011})%
  \bibAnnoteFile{NoStop}{Radisavljevic2011}%
\bibitem{Lembke2012}%
  \BibitemOpen
  \bibfield{author}{%
  \bibinfo {author} {\bibfnamefont{D.}~\bibnamefont{Lembke}}\ and\ \bibinfo
  {author} {\bibfnamefont{A.}~\bibnamefont{Kis}},\ }%
  \bibfield{journal}{%
  \Doi{10.1021/nn303772b}{\bibinfo {journal} {ACS Nano}}\ }%
  \textbf{\bibinfo {volume} {6}},\ \bibinfo {pages} {10070} (\bibinfo {month}
  {11}\ \bibinfo {year} {2012})%
  \bibAnnoteFile{NoStop}{Lembke2012}%
\bibitem{Wang2012}%
  \BibitemOpen
  \bibfield{author}{%
  \bibinfo {author} {\bibfnamefont{H.}~\bibnamefont{Wang}}, \bibinfo {author}
  {\bibfnamefont{L.}~\bibnamefont{Yu}}, \bibinfo {author}
  {\bibfnamefont{Y.-H.}\ \bibnamefont{Lee}}, \bibinfo {author}
  {\bibfnamefont{Y.}~\bibnamefont{Shi}}, \bibinfo {author}
  {\bibfnamefont{A.}~\bibnamefont{Hsu}}, \bibinfo {author}
  {\bibfnamefont{M.~L.}\ \bibnamefont{Chin}}, \bibinfo {author}
  {\bibfnamefont{L.-J.}\ \bibnamefont{Li}}, \bibinfo {author}
  {\bibfnamefont{M.}~\bibnamefont{Dubey}}, \bibinfo {author}
  {\bibfnamefont{J.}~\bibnamefont{Kong}},\ and\ \bibinfo {author}
  {\bibfnamefont{T.}~\bibnamefont{Palacios}},\ }%
  \bibfield{journal}{%
  \bibinfo {journal} {Nano Lett.}\ }%
  \textbf{\bibinfo {volume} {12}},\ \bibinfo {pages} {4674} (\bibinfo {year}
  {2012})%
  \bibAnnoteFile{NoStop}{Wang2012}%
\bibitem{Bollinger2001a}%
  \BibitemOpen
  \bibfield{author}{%
  \bibinfo {author} {\bibfnamefont{M.~V.}\ \bibnamefont{Bollinger}}, \bibinfo
  {author} {\bibfnamefont{J.~V.}\ \bibnamefont{Lauritsen}}, \bibinfo {author}
  {\bibfnamefont{K.~W.}\ \bibnamefont{Jacobsen}}, \bibinfo {author}
  {\bibfnamefont{J.~K.}\ \bibnamefont{N\o~rskov}}, \bibinfo {author}
  {\bibfnamefont{S.}~\bibnamefont{Helveg}},\ and\ \bibinfo {author}
  {\bibfnamefont{F.}~\bibnamefont{Besenbacher}},\ }%
  \bibfield{journal}{%
  \bibinfo {journal} {Phys. Rev. Lett.}\ }%
  \textbf{\bibinfo {volume} {87}},\ \bibinfo {pages} {196803} (\bibinfo {year}
  {2001})%
  \bibAnnoteFile{NoStop}{Bollinger2001a}%
\bibitem{Pan2012a}%
  \BibitemOpen
  \bibfield{author}{%
  \bibinfo {author} {\bibfnamefont{H.}~\bibnamefont{Pan}}\ and\ \bibinfo
  {author} {\bibfnamefont{Y.-W.}\ \bibnamefont{Zhang}},\ }%
  \bibfield{journal}{%
  \bibinfo {journal} {J. Mater. Chem.}\ }%
  \textbf{\bibinfo {volume} {22}},\ \bibinfo {pages} {7280} (\bibinfo {year}
  {2012})%
  \bibAnnoteFile{NoStop}{Pan2012a}%
\bibitem{Erdogan:2012aa}%
  \BibitemOpen
  \bibfield{author}{%
  \bibinfo {author} {\bibfnamefont{E.}~\bibnamefont{Erdogan}}, \bibinfo
  {author} {\bibfnamefont{I.~H.}\ \bibnamefont{Popov}}, \bibinfo {author}
  {\bibfnamefont{A.~N.}\ \bibnamefont{Enyashin}},\ and\ \bibinfo {author}
  {\bibfnamefont{G.}~\bibnamefont{Seifert}},\ }%
  \bibfield{journal}{%
  \Doi{10.1140/epjb/e2011-20456-7}{\bibinfo {journal} {Eur. Phys. J. B}}\ }%
  \textbf{\bibinfo {volume} {85}},\ \bibinfo {pages} {33} (\bibinfo {year}
  {2012})%
  \bibAnnoteFile{NoStop}{Erdogan:2012aa}%
\bibitem{Pavlovic2015b}%
  \BibitemOpen
  \bibfield{author}{%
  \bibinfo {author} {\bibfnamefont{S.}~\bibnamefont{Pavlovi\'{c}}}\ and\
  \bibinfo {author} {\bibfnamefont{F.~M.}\ \bibnamefont{Peeters}},\ }%
  \bibfield{journal}{%
  \bibinfo {journal} {Phys. Rev. B}\ }%
  \textbf{\bibinfo {volume} {91}},\ \bibinfo {pages} {155410} (\bibinfo {year}
  {2015})%
  \bibAnnoteFile{NoStop}{Pavlovic2015b}%
\bibitem{Davelou2014a}%
  \BibitemOpen
  \bibfield{author}{%
  \bibinfo {author} {\bibfnamefont{D.}~\bibnamefont{Davelou}}, \bibinfo
  {author} {\bibfnamefont{G.}~\bibnamefont{Kopidakis}}, \bibinfo {author}
  {\bibfnamefont{G.}~\bibnamefont{Kioseoglou}},\ and\ \bibinfo {author}
  {\bibfnamefont{I.~N.}\ \bibnamefont{Remediakis}},\ }%
  \bibfield{journal}{%
  \bibinfo {journal} {Solid State Commun.}\ }%
  \textbf{\bibinfo {volume} {192}},\ \bibinfo {pages} {42} (\bibinfo {year}
  {2014})%
  \bibAnnoteFile{NoStop}{Davelou2014a}%
\bibitem{Peterfalvi2015}%
  \BibitemOpen
  \bibfield{author}{%
  \bibinfo {author} {\bibfnamefont{C.~G.}\ \bibnamefont{P\'{e}terfalvi}},
  \bibinfo {author} {\bibfnamefont{A.}~\bibnamefont{Korm\'{a}nyos}},\ and\
  \bibinfo {author} {\bibfnamefont{G.}~\bibnamefont{Burkard}},\ }%
  \bibfield{journal}{%
  \bibinfo {journal} {arXiv:1509.00184}}%
   (\bibinfo {year} {2015})%
  \bibAnnoteFile{NoStop}{Peterfalvi2015}%
\bibitem{Gong:2014aa}%
  \BibitemOpen
  \bibfield{author}{%
  \bibinfo {author} {\bibfnamefont{Y.}~\bibnamefont{Gong}}, \bibinfo {author}
  {\bibfnamefont{J.}~\bibnamefont{Lin}}, \bibinfo {author}
  {\bibfnamefont{X.}~\bibnamefont{Wang}}, \bibinfo {author}
  {\bibfnamefont{G.}~\bibnamefont{Shi}}, \bibinfo {author}
  {\bibfnamefont{S.}~\bibnamefont{Lei}}, \bibinfo {author}
  {\bibfnamefont{Z.}~\bibnamefont{Lin}}, \bibinfo {author}
  {\bibfnamefont{X.}~\bibnamefont{Zou}}, \bibinfo {author}
  {\bibfnamefont{G.}~\bibnamefont{Ye}}, \bibinfo {author}
  {\bibfnamefont{R.}~\bibnamefont{Vajtai}}, \bibinfo {author}
  {\bibfnamefont{B.~I.}\ \bibnamefont{Yakobson}}, \bibinfo {author}
  {\bibfnamefont{H.}~\bibnamefont{Terrones}}, \bibinfo {author}
  {\bibfnamefont{M.}~\bibnamefont{Terrones}}, \bibinfo {author}
  {\bibfnamefont{B.~K.}\ \bibnamefont{Tay}}, \bibinfo {author}
  {\bibfnamefont{J.}~\bibnamefont{Lou}}, \bibinfo {author}
  {\bibfnamefont{S.~T.}\ \bibnamefont{Pantelides}}, \bibinfo {author}
  {\bibfnamefont{Z.}~\bibnamefont{Liu}}, \bibinfo {author}
  {\bibfnamefont{W.}~\bibnamefont{Zhou}},\ and\ \bibinfo {author}
  {\bibfnamefont{P.~M.}\ \bibnamefont{Ajayan}},\ }%
  \bibfield{journal}{%
  \bibinfo {journal} {Nat. Mater.}\ }%
  \textbf{\bibinfo {volume} {13}},\ \bibinfo {pages} {1135} (\bibinfo {month}
  {12}\ \bibinfo {year} {2014})%
  \bibAnnoteFile{NoStop}{Gong:2014aa}%
\bibitem{Chen:2015aa}%
  \BibitemOpen
  \bibfield{author}{%
  \bibinfo {author} {\bibfnamefont{K.}~\bibnamefont{Chen}}, \bibinfo {author}
  {\bibfnamefont{X.}~\bibnamefont{Wan}}, \bibinfo {author}
  {\bibfnamefont{J.}~\bibnamefont{Wen}}, \bibinfo {author}
  {\bibfnamefont{W.}~\bibnamefont{Xie}}, \bibinfo {author}
  {\bibfnamefont{Z.}~\bibnamefont{Kang}}, \bibinfo {author}
  {\bibfnamefont{X.}~\bibnamefont{Zeng}}, \bibinfo {author}
  {\bibfnamefont{H.}~\bibnamefont{Chen}},\ and\ \bibinfo {author}
  {\bibfnamefont{J.-B.}\ \bibnamefont{Xu}},\ }%
  \bibfield{journal}{%
  \Doi{10.1021/acsnano.5b03188}{\bibinfo {journal} {ACS Nano}}\ }%
  \textbf{\bibinfo {volume} {9}},\ \bibinfo {pages} {9868} (\bibinfo {month}
  {09}\ \bibinfo {year} {2015})%
  \bibAnnoteFile{NoStop}{Chen:2015aa}%
\bibitem{Hasan2010}%
  \BibitemOpen
  \bibfield{author}{%
  \bibinfo {author} {\bibfnamefont{M.~Z.}\ \bibnamefont{Hasan}}\ and\ \bibinfo
  {author} {\bibfnamefont{C.~L.}\ \bibnamefont{Kane}},\ }%
  \bibfield{journal}{%
  \bibinfo {journal} {Rev. Mod. Phys.}\ }%
  \textbf{\bibinfo {volume} {82}},\ \bibinfo {pages} {3045} (\bibinfo {year}
  {2010})%
  \bibAnnoteFile{NoStop}{Hasan2010}%
\bibitem{APLMat}%
  \BibitemOpen
  \enquote{\bibinfo {title} {Special topic on topological insulators},}\
  \bibinfo {howpublished} {APL Mat. {\bf3}, issue 8} (\bibinfo {year} {2015})%
  \bibAnnoteFile{NoStop}{APLMat}%
\bibitem{Teo2010}%
  \BibitemOpen
  \bibfield{author}{%
  \bibinfo {author} {\bibfnamefont{J.~C.~Y.}\ \bibnamefont{Teo}}\ and\ \bibinfo
  {author} {\bibfnamefont{C.~L.}\ \bibnamefont{Kane}},\ }%
  \bibfield{journal}{%
  \bibinfo {journal} {Phys. Rev. B}\ }%
  \textbf{\bibinfo {volume} {82}},\ \bibinfo {pages} {115120} (\bibinfo {year}
  {2010})%
  \bibAnnoteFile{NoStop}{Teo2010}%
\bibitem{Li2011}%
  \BibitemOpen
  \bibfield{author}{%
  \bibinfo {author} {\bibfnamefont{J.}~\bibnamefont{Li}}, \bibinfo {author}
  {\bibfnamefont{I.}~\bibnamefont{Martin}}, \bibinfo {author}
  {\bibfnamefont{M.}~\bibnamefont{B\"{u}ttiker}},\ and\ \bibinfo {author}
  {\bibfnamefont{A.~F.}\ \bibnamefont{Morpurgo}},\ }%
  \bibfield{journal}{%
  \bibinfo {journal} {Nat. Phys.}\ }%
  \textbf{\bibinfo {volume} {7}},\ \bibinfo {pages} {38} (\bibinfo {year}
  {2011})%
  \bibAnnoteFile{NoStop}{Li2011}%
\bibitem{Yao2009}%
  \BibitemOpen
  \bibfield{author}{%
  \bibinfo {author} {\bibfnamefont{W.}~\bibnamefont{Yao}}, \bibinfo {author}
  {\bibfnamefont{S.~A.}\ \bibnamefont{Yang}},\ and\ \bibinfo {author}
  {\bibfnamefont{Q.}~\bibnamefont{Niu}},\ }%
  \bibfield{journal}{%
  \bibinfo {journal} {Phys. Rev. Lett.}\ }%
  \textbf{\bibinfo {volume} {102}},\ \bibinfo {pages} {096801} (\bibinfo {year}
  {2009})%
  \bibAnnoteFile{NoStop}{Yao2009}%
\bibitem{Li2010}%
  \BibitemOpen
  \bibfield{author}{%
  \bibinfo {author} {\bibfnamefont{J.}~\bibnamefont{Li}}, \bibinfo {author}
  {\bibfnamefont{A.~F.}\ \bibnamefont{Morpurgo}}, \bibinfo {author}
  {\bibfnamefont{M.}~\bibnamefont{B\"{u}ttiker}},\ and\ \bibinfo {author}
  {\bibfnamefont{I.}~\bibnamefont{Martin}},\ }%
  \bibfield{journal}{%
  \bibinfo {journal} {Phys. Rev. B}\ }%
  \textbf{\bibinfo {volume} {82}},\ \bibinfo {pages} {245404} (\bibinfo {year}
  {2010})%
  \bibAnnoteFile{NoStop}{Li2010}%
\bibitem{Liu2013a}%
  \BibitemOpen
  \bibfield{author}{%
  \bibinfo {author} {\bibfnamefont{G.~B.}\ \bibnamefont{Liu}}, \bibinfo
  {author} {\bibfnamefont{W.~Y.}\ \bibnamefont{Shan}}, \bibinfo {author}
  {\bibfnamefont{Y.}~\bibnamefont{Yao}}, \bibinfo {author}
  {\bibfnamefont{W.}~\bibnamefont{Yao}},\ and\ \bibinfo {author}
  {\bibfnamefont{D.}~\bibnamefont{Xiao}},\ }%
  \bibfield{journal}{%
  \bibinfo {journal} {Phys. Rev. B}\ }%
  \textbf{\bibinfo {volume} {88}},\ \bibinfo {pages} {085433} (\bibinfo {year}
  {2013})%
  \bibAnnoteFile{NoStop}{Liu2013a}%
\bibitem{Kormanyos2013a}%
  \BibitemOpen
  \bibfield{author}{%
  \bibinfo {author} {\bibfnamefont{A.}~\bibnamefont{Korm\'{a}nyos}}, \bibinfo
  {author} {\bibfnamefont{V.}~\bibnamefont{Z\'{o}lyomi}}, \bibinfo {author}
  {\bibfnamefont{N.~D.}\ \bibnamefont{Drummond}}, \bibinfo {author}
  {\bibfnamefont{P.}~\bibnamefont{Rakyta}}, \bibinfo {author}
  {\bibfnamefont{G.}~\bibnamefont{Burkard}},\ and\ \bibinfo {author}
  {\bibfnamefont{V.~I.}\ \bibnamefont{Fal'ko}},\ }%
  \bibfield{journal}{%
  \bibinfo {journal} {Phys. Rev. B}\ }%
  \textbf{\bibinfo {volume} {88}},\ \bibinfo {pages} {045416} (\bibinfo {year}
  {2013})%
  \bibAnnoteFile{NoStop}{Kormanyos2013a}%
\bibitem{Rostami2013a}%
  \BibitemOpen
  \bibfield{author}{%
  \bibinfo {author} {\bibfnamefont{H.}~\bibnamefont{Rostami}}, \bibinfo
  {author} {\bibfnamefont{A.~G.}\ \bibnamefont{Moghaddam}},\ and\ \bibinfo
  {author} {\bibfnamefont{R.}~\bibnamefont{Asgari}},\ }%
  \bibfield{journal}{%
  \bibinfo {journal} {Phys. Rev. B}\ }%
  \textbf{\bibinfo {volume} {88}},\ \bibinfo {pages} {085440} (\bibinfo {year}
  {2013})%
  \bibAnnoteFile{NoStop}{Rostami2013a}%
\bibitem{Kormanyos2015}%
  \BibitemOpen
  \bibfield{author}{%
  \bibinfo {author} {\bibfnamefont{A.}~\bibnamefont{Korm\'{a}nyos}}, \bibinfo
  {author} {\bibfnamefont{G.}~\bibnamefont{Burkard}}, \bibinfo {author}
  {\bibfnamefont{M.}~\bibnamefont{Gmitra}}, \bibinfo {author}
  {\bibfnamefont{J.}~\bibnamefont{Fabian}}, \bibinfo {author}
  {\bibfnamefont{V.}~\bibnamefont{Z\'{o}lyomi}}, \bibinfo {author}
  {\bibfnamefont{N.~D.}\ \bibnamefont{Drummond}},\ and\ \bibinfo {author}
  {\bibfnamefont{V.}~\bibnamefont{Fal'ko}},\ }%
  \bibfield{journal}{%
  \bibinfo {journal} {2D Mater.}\ }%
  \textbf{\bibinfo {volume} {2}},\ \bibinfo {pages} {022001} (\bibinfo {year}
  {2015})%
  \bibAnnoteFile{NoStop}{Kormanyos2015}%
\bibitem{Kosmider:2013qf}%
  \BibitemOpen
  \bibfield{author}{%
  \bibinfo {author} {\bibfnamefont{K.}~\bibnamefont{Ko{\'s}mider}}, \bibinfo
  {author} {\bibfnamefont{J.~W.}\ \bibnamefont{Gonz{\'a}lez}},\ and\ \bibinfo
  {author} {\bibfnamefont{J.}~\bibnamefont{Fern{\'a}ndez-Rossier}},\ }%
  \bibfield{journal}{%
  \bibinfo {journal} {Phys. Rev. B}\ }%
  \textbf{\bibinfo {volume} {88}},\ \bibinfo {pages} {245436} (\bibinfo {month}
  {12}\ \bibinfo {year} {2013})%
  \bibAnnoteFile{NoStop}{Kosmider:2013qf}%
\bibitem{Kormanyos:2014fk}%
  \BibitemOpen
  \bibfield{author}{%
  \bibinfo {author} {\bibfnamefont{A.}~\bibnamefont{Korm{\'a}nyos}}, \bibinfo
  {author} {\bibfnamefont{V.}~\bibnamefont{Z{\'o}lyomi}}, \bibinfo {author}
  {\bibfnamefont{N.~D.}\ \bibnamefont{Drummond}},\ and\ \bibinfo {author}
  {\bibfnamefont{G.}~\bibnamefont{Burkard}},\ }%
  \bibfield{journal}{%
  \bibinfo {journal} {Phys. Rev. X}\ }%
  \textbf{\bibinfo {volume} {4}},\ \bibinfo {pages} {011034} (\bibinfo {month}
  {03}\ \bibinfo {year} {2014})%
  \bibAnnoteFile{NoStop}{Kormanyos:2014fk}%
\bibitem{Note1}%
  \BibitemOpen
  \bibinfo {note} {We do not consider parameters $\alpha $ and $\beta $ with
  the same sign since in these cases there is not a real gap separating the
  bands.}%
  \bibAnnoteFile{Stop}{Note1}%
\bibitem{BHZ2006}%
  \BibitemOpen
  \bibfield{author}{%
  \bibinfo {author} {\bibfnamefont{B.~A.}\ \bibnamefont{Bernevig}}, \bibinfo
  {author} {\bibfnamefont{T.~L.}\ \bibnamefont{Hughes}},\ and\ \bibinfo
  {author} {\bibfnamefont{S.-C.}\ \bibnamefont{Zhang}},\ }%
  \bibfield{journal}{%
  \bibinfo {journal} {Science}\ }%
  \textbf{\bibinfo {volume} {314}},\ \bibinfo {pages} {1757} (\bibinfo {year}
  {2006})%
  \bibAnnoteFile{NoStop}{BHZ2006}%
\bibitem{Gorbachev2014}%
  \BibitemOpen
  \bibfield{author}{%
  \bibinfo {author} {\bibfnamefont{R.~V.}\ \bibnamefont{Gorbachev}}, \bibinfo
  {author} {\bibfnamefont{J.~C.~W.}\ \bibnamefont{Song}}, \bibinfo {author}
  {\bibfnamefont{G.~L.}\ \bibnamefont{Yu}}, \bibinfo {author}
  {\bibfnamefont{A.~V.}\ \bibnamefont{Kretinin}}, \bibinfo {author}
  {\bibfnamefont{F.}~\bibnamefont{Withers}}, \bibinfo {author}
  {\bibfnamefont{Y.}~\bibnamefont{Cao}}, \bibinfo {author}
  {\bibfnamefont{A.}~\bibnamefont{Mishchenko}}, \bibinfo {author}
  {\bibfnamefont{I.~V.}\ \bibnamefont{Grigorieva}}, \bibinfo {author}
  {\bibfnamefont{K.~S.}\ \bibnamefont{Novoselov}}, \bibinfo {author}
  {\bibfnamefont{L.~S.}\ \bibnamefont{Levitov}},\ and\ \bibinfo {author}
  {\bibfnamefont{A.~K.}\ \bibnamefont{Geim}},\ }%
  \bibfield{journal}{%
  \bibinfo {journal} {Science}\ }%
  \textbf{\bibinfo {volume} {346}},\ \bibinfo {pages} {448} (\bibinfo {month}
  {10}\ \bibinfo {year} {2014})%
  \bibAnnoteFile{NoStop}{Gorbachev2014}%
\bibitem{Shimazaki2015}%
  \BibitemOpen
  \bibfield{author}{%
  \bibinfo {author} {\bibfnamefont{Y.}~\bibnamefont{Shimazaki}}, \bibinfo
  {author} {\bibfnamefont{M.}~\bibnamefont{Yamamoto}}, \bibinfo {author}
  {\bibfnamefont{I.~V.}\ \bibnamefont{Borzenets}}, \bibinfo {author}
  {\bibfnamefont{K.}~\bibnamefont{Watanabe}}, \bibinfo {author}
  {\bibfnamefont{T.}~\bibnamefont{Taniguchi}},\ and\ \bibinfo {author}
  {\bibfnamefont{S.}~\bibnamefont{Tarucha}},\ }%
  \bibfield{journal}{%
  \bibinfo {journal} {arXiv:1501.04776}}%
   (\bibinfo {year} {2015})%
  \bibAnnoteFile{NoStop}{Shimazaki2015}%
\bibitem{Duan2014}%
  \BibitemOpen
  \bibfield{author}{%
  \bibinfo {author} {\bibfnamefont{X.}~\bibnamefont{Duan}}, \bibinfo {author}
  {\bibfnamefont{C.}~\bibnamefont{Wang}}, \bibinfo {author}
  {\bibfnamefont{J.~C.}\ \bibnamefont{Shaw}}, \bibinfo {author}
  {\bibfnamefont{R.}~\bibnamefont{Cheng}}, \bibinfo {author}
  {\bibfnamefont{Y.}~\bibnamefont{Chen}}, \bibinfo {author}
  {\bibfnamefont{H.}~\bibnamefont{Li}}, \bibinfo {author}
  {\bibfnamefont{X.}~\bibnamefont{Wu}}, \bibinfo {author}
  {\bibfnamefont{Y.}~\bibnamefont{Tang}}, \bibinfo {author}
  {\bibfnamefont{Q.}~\bibnamefont{Zhang}}, \bibinfo {author}
  {\bibfnamefont{A.}~\bibnamefont{Pan}}, \bibinfo {author}
  {\bibfnamefont{J.}~\bibnamefont{Jiang}}, \bibinfo {author}
  {\bibfnamefont{R.}~\bibnamefont{Yu}}, \bibinfo {author}
  {\bibfnamefont{Y.}~\bibnamefont{Huang}},\ and\ \bibinfo {author}
  {\bibfnamefont{X.}~\bibnamefont{Duan}},\ }%
  \bibfield{journal}{%
  \bibinfo {journal} {Nat. Nanotechnol.}\ }%
  \textbf{\bibinfo {volume} {9}},\ \bibinfo {pages} {1024} (\bibinfo {year}
  {2014})%
  \bibAnnoteFile{NoStop}{Duan2014}%
\bibitem{Marshall2014}%
  \BibitemOpen
  \bibfield{author}{%
  \bibinfo {author} {\bibfnamefont{M.~S.~J.}\ \bibnamefont{Marshall}}\ and\
  \bibinfo {author} {\bibfnamefont{M.~R.}\ \bibnamefont{Castell}},\ }%
  \bibfield{journal}{%
  \bibinfo {journal} {Chem. Soc. Rev.}\ }%
  \textbf{\bibinfo {volume} {43}},\ \bibinfo {pages} {2226} (\bibinfo {year}
  {2014})%
  \bibAnnoteFile{NoStop}{Marshall2014}%
\bibitem{Liu2014}%
  \BibitemOpen
  \bibfield{author}{%
  \bibinfo {author} {\bibfnamefont{Z.}~\bibnamefont{Liu}}, \bibinfo {author}
  {\bibfnamefont{M.}~\bibnamefont{Amani}}, \bibinfo {author}
  {\bibfnamefont{S.}~\bibnamefont{Najmaei}}, \bibinfo {author}
  {\bibfnamefont{Q.}~\bibnamefont{Xu}}, \bibinfo {author}
  {\bibfnamefont{X.}~\bibnamefont{Zou}}, \bibinfo {author}
  {\bibfnamefont{W.}~\bibnamefont{Zhou}}, \bibinfo {author}
  {\bibfnamefont{T.}~\bibnamefont{Yu}}, \bibinfo {author}
  {\bibfnamefont{C.}~\bibnamefont{Qiu}}, \bibinfo {author}
  {\bibfnamefont{A.~G.}\ \bibnamefont{Birdwell}}, \bibinfo {author}
  {\bibfnamefont{F.~J.}\ \bibnamefont{Crowne}}, \bibinfo {author}
  {\bibfnamefont{R.}~\bibnamefont{Vajtai}}, \bibinfo {author}
  {\bibfnamefont{B.~I.}\ \bibnamefont{Yakobson}}, \bibinfo {author}
  {\bibfnamefont{Z.}~\bibnamefont{Xia}}, \bibinfo {author}
  {\bibfnamefont{M.}~\bibnamefont{Dubey}}, \bibinfo {author}
  {\bibfnamefont{P.~M.}\ \bibnamefont{Ajayan}},\ and\ \bibinfo {author}
  {\bibfnamefont{J.}~\bibnamefont{Lou}},\ }%
  \bibfield{journal}{%
  \bibinfo {journal} {Nat. Commun.}\ }%
  \textbf{\bibinfo {volume} {5}},\ \bibinfo {pages} {5246} (\bibinfo {year}
  {2014})%
  \bibAnnoteFile{NoStop}{Liu2014}%
\bibitem{Bao2015}%
  \BibitemOpen
  \bibfield{author}{%
  \bibinfo {author} {\bibfnamefont{W.}~\bibnamefont{Bao}}, \bibinfo {author}
  {\bibfnamefont{N.~J.}\ \bibnamefont{Borys}}, \bibinfo {author}
  {\bibfnamefont{C.}~\bibnamefont{Ko}}, \bibinfo {author}
  {\bibfnamefont{J.}~\bibnamefont{Suh}}, \bibinfo {author}
  {\bibfnamefont{W.}~\bibnamefont{Fan}}, \bibinfo {author}
  {\bibfnamefont{A.}~\bibnamefont{Thron}}, \bibinfo {author}
  {\bibfnamefont{Y.}~\bibnamefont{Zhang}}, \bibinfo {author}
  {\bibfnamefont{A.}~\bibnamefont{Buyanin}}, \bibinfo {author}
  {\bibfnamefont{J.}~\bibnamefont{Zhang}}, \bibinfo {author}
  {\bibfnamefont{S.}~\bibnamefont{Cabrini}}, \bibinfo {author}
  {\bibfnamefont{P.~D.}\ \bibnamefont{Ashby}}, \bibinfo {author}
  {\bibfnamefont{A.}~\bibnamefont{Weber-Bargioni}}, \bibinfo {author}
  {\bibfnamefont{S.}~\bibnamefont{Tongay}}, \bibinfo {author}
  {\bibfnamefont{S.}~\bibnamefont{Aloni}}, \bibinfo {author}
  {\bibfnamefont{D.~F.}\ \bibnamefont{Ogletree}}, \bibinfo {author}
  {\bibfnamefont{J.}~\bibnamefont{Wu}}, \bibinfo {author}
  {\bibfnamefont{M.~B.}\ \bibnamefont{Salmeron}},\ and\ \bibinfo {author}
  {\bibfnamefont{P.~J.}\ \bibnamefont{Schuck}},\ }%
  \bibfield{journal}{%
  \bibinfo {journal} {Nat. Commun.}\ }%
  \textbf{\bibinfo {volume} {6}},\ \bibinfo {pages} {7993} (\bibinfo {year}
  {2015})%
  \bibAnnoteFile{NoStop}{Bao2015}%
\end{thebibliography}
%

\end{document}